\RequirePackage{fix-cm}
\documentclass[twocolumn]{svjour3}          % twocolumn
\smartqed  % flush right qed marks, e.g. at end of proof
\usepackage{graphicx}
\usepackage{natbib}
\usepackage{amsmath}
\usepackage{txfonts}
\usepackage{textcomp}
\usepackage{tabularx}
\journalname{IAG Symposia Series}
\begin{document}

\title{Testing special relativity with geodetic VLBI}


\author{Oleg Titov          \and
        Hana Kr{\'a}sn{\'a} %etc.
}


\institute{O. Titov \at
              Geoscience Australia, PO Box 378, Canberra, 2601, Australia\\
              \email{Oleg.Titov@ga.gov.au}           %  \\
           \and
              H. Kr{\'a}sn{\'a} \at
           1. Technische Universit{\"a}t Wien, Vienna, Austria \\
           2. Astronomical Institute, Czech Academy of Sciences, Prague, Czech Republic
}

\date{Received: date / Accepted: date}

\maketitle

\begin{abstract}
Geodetic Very Long Baseline Interferometry (VLBI) measures the group delay in the barycentric reference frame. As the Earth is orbiting around the Solar system barycentre with the velocity $V$ of 30~km/s, VLBI proves to be a handy tool to detect the subtle effects of the special and general relativity theory with a magnitude of $(V/\textrm{c})^2$. The theoretical correction for the second order terms reaches up to 300~ps, and it is implemented in the geodetic VLBI group delay model. The total contribution of the second order terms splits into two effects - the variation of the Earth scale, and the deflection of the apparent position of the radio source.
The Robertson-Mansouri-Sexl (RMS) generalization of the Lorenz transformation is used for many modern tests of the special relativity theory. We develop an alteration of the RMS formalism to probe the Lorenz invariance with the geodetic VLBI data. The kinematic approach implies three parameters (as a function of the moving reference frame velocity) and the standard Einstein synchronisation. A generalised relativistic model of geodetic VLBI data includes all three parameters that could be estimated. Though, since the modern laboratory Michelson-Morley and Kennedy-Thorndike experiments are more accurate than VLBI technique, the presented equations may be used to test the VLBI group delay model itself.

\end{abstract}

\section{Introduction}
The Very Long Baseline Interferometry (VLBI) technique measures time delay - the difference between times of the signal arrival on two radio telescopes separated by a long baseline. All measurements are referred to the Solar system barycentre, which moves around the Sun with an orbital velocity about 30~km/s. This makes the Earth a natural flying platform and VLBI a very effective tool to detect a tiny effect of special relativity. Each baseline of thousand kilometres length may serve as a ``flying rod", which is traditionally used for theoretical calculation. Precision of each single group delay is about 10~mm and since many observations are collected over a long period of time (20~years or more) the estimate of the time dilation effect will be very accurate.

Geodetic VLBI has been used to test general relativity theory either in the frame of the Parameterized Post-Newtonian (PPN) formalism or the Standard-Model Extension (SME) (e.g., \citet{Robertson84}, \citet{Shapiro04}, \citet{Lambert09}, or \citet{LePoncinL16}). However, it has not been considered for testing special relativity in spite of its interferometric nature directly linked to the Michelson-Morley and Kennedy-Thorndike interferometers yet. In this paper we show a possible application of the geodetic VLBI to this experimental work.

\begin{figure}[tbp]
          \includegraphics[trim=0 0 0 0,clip, width=\hsize]{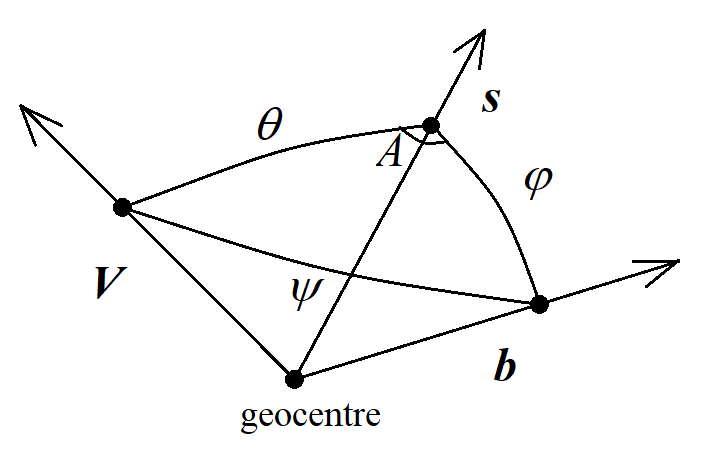}
  \caption{Schematic view of the introduced angles with vectors placed in the geocentre.}
     \label{fig_triangle}
\end{figure}

The conventional group delay $\tau_{g}$ model to approximate the observed VLBI data is given
by~\citet[chap.~11]{iers10} as
\begin{equation}\label{groupdelay_gcrs}
\tau_{g}=\frac{-\frac{(\boldsymbol{b}\cdot\boldsymbol{s})}{\textrm{c}}\Big(1-\frac{2GM}{\textrm{c}^{2} R} -\frac{|\boldsymbol{V}|^{2}}{2\textrm{c}^{2}}-\frac{(\boldsymbol{V}\cdot\boldsymbol{w}_{2})}{\textrm{c}^2}\Big) -\frac{(\boldsymbol{b}\cdot\boldsymbol{V})}{\textrm{c}^{2}}\Big(1+\frac{(\boldsymbol{s}\cdot\boldsymbol{V})}{2\textrm{c}}\Big)}
    {1+\frac{\boldsymbol{s}\cdot(\boldsymbol{V}+\boldsymbol{w}_{2})}{\textrm{c}}}
\end{equation}
where $\boldsymbol{b}$ is the vector of baseline $\boldsymbol{b} = \boldsymbol{r_2}-\boldsymbol{r_1}$, $\boldsymbol{s}$ is the barycentric unit vector of radio source, $\boldsymbol{V}$ is the barycentric velocity of the geocentre, $\boldsymbol{w}_{2}$ is the geocentric velocity of the second station, \textrm{c} is the speed of light, \textrm{G} is the gravitational constant, $M$ is the mass of the Sun, and $R$ is the geocentric distance to the Sun.\\
The term $\frac{2GM}{\textrm{c}^{2} R}$ is related to the general relativity effect and we won't focus on it in this note. The impact of $\boldsymbol{w}_{2}$ is small and may be ignored for the sake of simplicity. After these alterations, Equation~(\ref{groupdelay_gcrs}) is given by
\begin{equation}\label{groupdelay_gcrs_simple}
\tau_{g}=\frac{-\frac{(\boldsymbol{b}\cdot\boldsymbol{s})}{\textrm{c}}\Big(1 -\frac{|\boldsymbol{V}|^{2}}{2\textrm{c}^{2}}\Big) -\frac{1}{\textrm{c}^{2}}(\boldsymbol{b}\cdot\boldsymbol{V}) \Big(1+\frac{(\boldsymbol{s}\cdot\boldsymbol{V})}{2\textrm{c}}\Big)}
    {1+\frac{1}{\textrm{c}}(\boldsymbol{s}\cdot\boldsymbol{V})} .
\end{equation}
Using the Taylor series expansion $(1+x)^{-1} = 1-x+x^2$
for $\Big(1+\frac{(\boldsymbol{s}\cdot\boldsymbol{V})}{\textrm{c}}\Big)^{-1}$ and noting the $\frac{|\boldsymbol{V}|^2}{\textrm{c}^2}$ terms only, Equation~(\ref{groupdelay_gcrs_simple}) reduces to
\begin{equation}\label{groupdelay_gcrs_simple_v2c3}
\begin{aligned}
\tau_{g}=&\frac{(\boldsymbol{b}\cdot\boldsymbol{s})}{\textrm{c}}
\frac{|\boldsymbol{V}|^{2}}{2\textrm{c}^{2}} - \frac{1}{\textrm{c}^{2}} (\boldsymbol{b}\cdot\boldsymbol{V})
\frac{(\boldsymbol{s}\cdot\boldsymbol{V})}{2\textrm{c}} -\\
&-\frac{(\boldsymbol{b}\cdot\boldsymbol{s})(\boldsymbol{s}\cdot\boldsymbol{V})^2 }{\textrm{c}^3} +
\frac{1}{\textrm{c}^{2}} (\boldsymbol{b}\cdot\boldsymbol{V}) \frac{(\boldsymbol{s}\cdot\boldsymbol{V})}{\textrm{c}} =\\
=&\frac{(\boldsymbol{b}\cdot\boldsymbol{s})|\boldsymbol{V}|^{2}}{2\textrm{c}^{3}} -
\frac{(\boldsymbol{b}\cdot\boldsymbol{s})(\boldsymbol{s}\cdot\boldsymbol{V})^2}{2\textrm{c}^{3}} \\
&-\frac{(\boldsymbol{b}\cdot\boldsymbol{V})(\boldsymbol{s}\cdot\boldsymbol{V})}{2\textrm{c}^{3}} +
\frac{(\boldsymbol{b}\cdot\boldsymbol{V})(\boldsymbol{s}\cdot\boldsymbol{V})}{\textrm{c}^{3}} -
\frac{(\boldsymbol{b}\cdot\boldsymbol{s})(\boldsymbol{s}\cdot\boldsymbol{V})^2}{2\textrm{c}^{3}} = \\
=&\frac{(\boldsymbol{b}\cdot\boldsymbol{s})\Big(|\boldsymbol{V}|^{2}-(\boldsymbol{s}\cdot\boldsymbol{V})^2\Big)}{2\textrm{c}^{3}} +
\frac{(\boldsymbol{s}\cdot\boldsymbol{V})\Big((\boldsymbol{b}\cdot\boldsymbol{V})-(\boldsymbol{b}\cdot\boldsymbol{s})(\boldsymbol{s}\cdot\boldsymbol{V})\Big)}{2\textrm{c}^{3}} .
\end{aligned}
\end{equation}
In Fig.~\ref{fig_triangle} we introduce the following angles
$|\boldsymbol{V}| \cos\theta = (\boldsymbol{s}\cdot\boldsymbol{V})$,
$|\boldsymbol{b}| \cos\varphi = (\boldsymbol{b}\cdot\boldsymbol{s})$ and
$|\boldsymbol{b}||\boldsymbol{V}| \cos\psi = (\boldsymbol{b}\cdot\boldsymbol{V})$, and from the equation of spherical trigonometry we get $\cos\psi = \cos\theta \cos\varphi + \sin\theta \sin\varphi \cos A$. After applying the substitution
\begin{equation}\label{trigfunc1}
\begin{aligned}
(\boldsymbol{b}\cdot\boldsymbol{s})\Big(|\boldsymbol{V}|^{2}-(\boldsymbol{s}\cdot\boldsymbol{V})^2\Big) =
&|\boldsymbol{b}| \cos\varphi \Big(|\boldsymbol{V}|^2-|\boldsymbol{V}|^2 \cos^2\theta\Big) =\\
=&|\boldsymbol{b}| |\boldsymbol{V}|^2 \cos\varphi \sin^2\theta
\end{aligned}
\end{equation}
and
\begin{equation}\label{trigfunc2}
\begin{aligned}
&(\boldsymbol{s}\cdot\boldsymbol{V})\Big((\boldsymbol{b}\cdot\boldsymbol{V})-(\boldsymbol{b}\cdot\boldsymbol{s})(\boldsymbol{s}\cdot\boldsymbol{V})\Big) = \\
&=|\boldsymbol{V}| \cos\theta \Big( |\boldsymbol{b}||\boldsymbol{V}| \cos\psi -  (|\boldsymbol{b}| \cos\varphi)   (|\boldsymbol{V}| \cos\theta)       \Big) = \\
&=|\boldsymbol{b}||\boldsymbol{V}|^2 \cos\theta \Big(\cos\theta \cos\varphi  +  \sin\theta \sin\varphi \cos A - \cos\theta \cos\varphi\Big)
\end{aligned}
\end{equation}
we get the Equation~(\ref{groupdelay_gcrs_simple_v2c3}) in the following form
\begin{equation}\label{groupdelay_sphtrig}
\tau_{g}= \frac{|\boldsymbol{b}||\boldsymbol{V}|^{2}}{2\textrm{c}^3} \cos\varphi \sin^2\theta +
             \frac{|\boldsymbol{b}||\boldsymbol{V}|^{2}}{2\textrm{c}^3} \sin\varphi \sin\theta\cos\theta\cos A .
\end{equation}
\\

The major term of the geometric delay is
\begin{equation}\label{tau_geom}
\tau_{g}=-\frac{(\boldsymbol{b}\cdot\boldsymbol{s})}{\textrm{c}} = -\frac{|\boldsymbol{b}|\cos\varphi}{\textrm{c}} .
\end{equation}
The components of the baseline vector $\boldsymbol{b}$ and the source position vector $\boldsymbol{s}$ can be estimated from a large set of data within an adjustment. The observational delay from a correlator is approximated by the theoretical delay (Equation~(\ref{groupdelay_gcrs})), and the difference between the observational and the theoretical delay is modelled as follows:
\begin{equation}\label{dtau}
\tau_{obs} - \tau_{calc} = \frac{\partial \tau}{\partial b} \Delta b + \frac{\partial \tau}{\partial s} \Delta s
\end{equation}
or, by applying Equation~(\ref{tau_geom}) one gets
\begin{equation}\label{dtau1}
\tau_{obs} - \tau_{calc} = -\Delta b\frac{1}{\textrm{c}} \cos\varphi + \Delta s\frac{|\boldsymbol{b}|}{\textrm{c}}\sin\varphi
\end{equation}
which means that the corrections to the baseline vector components are calculated with the partials proportional to $\cos\varphi$ and corrections to the source vector components need partials proportional to $\sin\varphi$.
Therefore, the first part of Equation~(\ref{groupdelay_sphtrig}) is a variation of the baseline vector (i.e., of the Earth scale) as it is proportional to the factor $(|\boldsymbol{b}|\cos{\varphi})$, and the second part is a variation of the source positions $(|\boldsymbol{b}|\sin{\varphi})$.

\section{Lorenz transformation}
\label{sec:LorenzTrafo}
There are many approaches to generalise the standard Lorenz transformation. The first known is the Robertson-Mansouri-Sexl (RMS) formalism \citep{Robertson49, Mansouri77a, Mansouri77b} that assumes a possibility of anisotropic speed of light, i.e., tantamount of a preferred reference frame existing. Due to some critical discussion \citep{Rybicki16}, we develop a new functional form of three functions $A(V), B(V),$ and $D(V)$ instead of the original RMS formalism. \\
Let's consider the Lorenz transformation in a general form between the preferred reference frame $S(\boldsymbol{x},t)$ and another reference frame $S'(\boldsymbol{x'},t')$ moving with a relative velocity $\boldsymbol{V}$
\begin{equation}\label{xt}
\begin{aligned}
\boldsymbol{x'} = & D\boldsymbol{x}-(D-B)\frac{(\boldsymbol{V}\boldsymbol{x})\boldsymbol{V}}{|\boldsymbol{V}|^2} - B\boldsymbol{V}t \\
t' = & A\Bigg(t - \frac{\boldsymbol{V}}{\textrm{c}^2}\boldsymbol{x}\Bigg) .
\end{aligned}
\end{equation}
For the sake of simplicity, we adopt the Einstein synchronisation here. In special relativity $A(V) = B(V) =\\ \bigg(\sqrt{1-\frac{|\boldsymbol{V}|^2}{\textrm{c}^2}}\bigg)^{-1} = \gamma$ and $D(V) = 1$. This form of generalisation is similar to those proposed by \citet{Will92} with some alterations.\\
The velocity transformation between frames $S'$ and $S$ may be derived from Equation~(\ref{xt}) as
$\boldsymbol{v'_x} = \frac{\boldsymbol{dx'}}{dt'}$ and $\boldsymbol{v_x} = \frac{\boldsymbol{dx}}{dt}$
\begin{equation}\label{dxdt}
\begin{aligned}
\boldsymbol{dx'} = & D\boldsymbol{dx}-(D-B)\frac{(\boldsymbol{V}\boldsymbol{dx})\boldsymbol{V}}{|\boldsymbol{V}|^2} - B\boldsymbol{V}dt \\
dt' = & A\Bigg(dt - \frac{\boldsymbol{V}}{\textrm{c}^2}\boldsymbol{dx}\Bigg)
\end{aligned}
\end{equation}
\begin{equation}\label{vx}
\begin{aligned}
\boldsymbol{v_x'} =& \frac{\boldsymbol{dx'}}{dt'} =
\frac{ D\frac{\boldsymbol{dx}}{dt}-(D-B)\frac{ \Big(\boldsymbol{V}\frac{\boldsymbol{dx}}{dt} \Big)\boldsymbol{V}}{|\boldsymbol{V}|^2} - B\boldsymbol{V}}
{A\bigg(1 - \frac{\boldsymbol{V}\frac{\boldsymbol{dx}}{dt}}{\textrm{c}^2}\bigg)} \\
 =& \frac{ D\boldsymbol{v_x}-(D-B)\frac{ (\boldsymbol{V}\boldsymbol{v_x} )\boldsymbol{V}}{|\boldsymbol{V}|^2} - B\boldsymbol{V}}
{A\bigg(1 - \frac{\boldsymbol{V}\boldsymbol{v_x}}{\textrm{c}^2}\bigg)} .
\end{aligned}
\end{equation}
For the propagation of light from an extragalactic radio source one has to assign the unit vector in the direction of the source apparent position (moving system $S'$) as $\boldsymbol{s'} = -\frac{\boldsymbol{v'_x}}{\textrm{c}}$, and in the preferred reference frame as $\boldsymbol{s} = -\frac{\boldsymbol{v_x}}{\textrm{c}}$. Therefore, the transformation between $\boldsymbol{s'}$ and $\boldsymbol{s}$ is given by
\begin{equation}\label{s}
\boldsymbol{s'} = \frac{ D\boldsymbol{s}-(D-B)\frac{(\boldsymbol{V}\boldsymbol{s})\boldsymbol{V}}{|\boldsymbol{V}|^2} +
\frac{B\boldsymbol{V}}{\textrm{c}}}
{A\Big(1 + \frac{(\boldsymbol{V}\boldsymbol{s})}{\textrm{c}}\Big)} .
\end{equation}
Now we apply the traditional expansion \citep{Mansouri77a, Will92}:
\begin{equation}\label{ABD}
\begin{aligned}
A(V) &= 1+\alpha \frac{|\boldsymbol{V}|^2}{\textrm{c}^2} + ... \\
B(V) &= 1+\beta \frac{|\boldsymbol{V}|^2}{\textrm{c}^2} + ... \\
D(V) &= 1+\delta \frac{|\boldsymbol{V}|^2}{\textrm{c}^2} + ...
\end{aligned}
\end{equation}
to the Equation~(\ref{s}) which gives
\begin{equation}\label{s_traditional}
\boldsymbol{s'} = \frac{ \Big(1+\delta \frac{|\boldsymbol{V}|^2}{\textrm{c}^2}\Big)\boldsymbol{s}+\Big(\beta-\delta\Big)\frac{(\boldsymbol{V}\boldsymbol{s})\boldsymbol{V}}{\textrm{c}^2} +
\Big(1+\beta \frac{|\boldsymbol{V}|^2}{\textrm{c}^2}\Big)\frac{\boldsymbol{V}}{\textrm{c}}}
{\Big(1+\alpha \frac{|\boldsymbol{V}|^2}{\textrm{c}^2}\Big)\Big(1 + \frac{(\boldsymbol{V}\boldsymbol{s})}{\textrm{c}}\Big)} .
\end{equation}
With the help of the Taylor series expansion $(1+x)^{-1} = 1-x+x^2$ and keeping the terms to $\frac{\boldsymbol{V}^2}{\textrm{c}^2}$ only, we obtain from Equation~(\ref{s_traditional})
\begin{equation}\label{s_traditional_tillV2c2}
\boldsymbol{s'} = \boldsymbol{s} +
\frac{\boldsymbol{V}-\boldsymbol{s}(\boldsymbol{V}\boldsymbol{s})}{\textrm{c}} +
\frac{\boldsymbol{s} \Big( |\boldsymbol{V}|^2 (\delta - \alpha) + (\boldsymbol{V}\boldsymbol{s})^2\Big)}{\textrm{c}^2}  +
\frac{(\beta-\delta-1)\boldsymbol{V}(\boldsymbol{V}\boldsymbol{s})}{\textrm{c}^2} .
\end{equation}
The second term $\sim\frac{\boldsymbol{V}}{\textrm{c}}$ represents the annual aberration. The third term (proportional to the vector $\boldsymbol{s}$ only) does not affect the apparent position of the celestial objects. It could be ignored by the traditional observational astrometric techniques but it is essential for the geodetic VLBI. Finally, the last term in the Equation~(\ref{s_traditional_tillV2c2}) describes the second order correction ($\sim\frac{|\boldsymbol{V}|^2}{\textrm{c}^2}$) in the radio source positions.\\
Now we convert the Equation~(\ref{s_traditional_tillV2c2}) to the geometric delay $\tau$
\begin{equation}\label{geomdel_s}
\begin{aligned}
\tau = &-\frac{(\boldsymbol{b}(\boldsymbol{s'}-\boldsymbol{s}))}{\textrm{c}} =
-\frac{(\boldsymbol{b}\boldsymbol{V})-(\boldsymbol{b}\boldsymbol{s})(\boldsymbol{V}\boldsymbol{s})}{\textrm{c}^2} \\ &-\frac{(\boldsymbol{b}\boldsymbol{s})(|\boldsymbol{V}|^2(\delta-\alpha)+(\boldsymbol{V}\boldsymbol{s})^2)}{\textrm{c}^3}-
\frac{(\beta-\delta-1)(\boldsymbol{b}\boldsymbol{V})(\boldsymbol{V}\boldsymbol{s})}{\textrm{c}^3} .
\end{aligned}
\end{equation}
Since in special relativity $\alpha = \frac{1}{2}, \beta = \frac{1}{2}$ and $\delta = 0$, the Equation~(\ref{geomdel_s}) may be presented as
\begin{equation}\label{geomdel_spec}
\begin{aligned}
\tau = &-\frac{(\boldsymbol{b}(\boldsymbol{s'}-\boldsymbol{s}))}{\textrm{c}} =
-\frac{(\boldsymbol{b}\boldsymbol{V})-(\boldsymbol{b}\boldsymbol{s})(\boldsymbol{V}\boldsymbol{s})}{\textrm{c}^2} \\ &-\frac{(\boldsymbol{b}\boldsymbol{s})\Big(-\frac{1}{2}|\boldsymbol{V}|^2+(\boldsymbol{V}\boldsymbol{s})^2\Big)}{\textrm{c}^3}+
\frac{(\boldsymbol{b}\boldsymbol{V})(\boldsymbol{V}\boldsymbol{s})}{2\textrm{c}^3} \\
=&-\frac{(\boldsymbol{b}\boldsymbol{V})-(\boldsymbol{b}\boldsymbol{s})(\boldsymbol{V}\boldsymbol{s})}{\textrm{c}^2} +
\frac{(\boldsymbol{b}\boldsymbol{s})\Big(|\boldsymbol{V}|^2-(\boldsymbol{V}\boldsymbol{s})^2\Big)}{2\textrm{c}^3}\\
&+\frac{(\boldsymbol{b}\boldsymbol{V})(\boldsymbol{V}\boldsymbol{s})}{2\textrm{c}^3}
-\frac{(\boldsymbol{b}\boldsymbol{s})(\boldsymbol{V}\boldsymbol{s})^2}{2\textrm{c}^3}
\end{aligned}
\end{equation}
where the terms $\sim\frac{|\boldsymbol{V}|^2}{\textrm{c}^2}$ coincide to Equation~(\ref{groupdelay_gcrs_simple_v2c3}).

It is essential to note that the sum of the two last terms in Equation~(\ref{geomdel_s})
\begin{equation}\label{geomdel_s_2terms}
\begin{aligned}
\Delta\tau = &\frac{(\boldsymbol{b}\boldsymbol{s})|\boldsymbol{V}|^2(\alpha-\delta)}{\textrm{c}^3}
-\frac{(\boldsymbol{b}\boldsymbol{s})(\boldsymbol{V}\boldsymbol{s})^2}{\textrm{c}^3}-
\frac{(\beta-\delta-1)(\boldsymbol{b}\boldsymbol{V})(\boldsymbol{V}\boldsymbol{s})}{\textrm{c}^3} \\
= & \frac{|\boldsymbol{b}||\boldsymbol{V}|^2(\alpha-\delta)}{\textrm{c}^3}\cos\varphi -
\frac{|\boldsymbol{b}||\boldsymbol{V}|^2}{\textrm{c}^3}\cos\varphi\cos^2\theta \\
&-\frac{|\boldsymbol{b}||\boldsymbol{V}|^2(\beta-\delta-1)}{\textrm{c}^3} (\cos\varphi \cos^2\theta - \sin\theta \cos\theta \sin\varphi \cos A)\\
= & \frac{|\boldsymbol{b}||\boldsymbol{V}|^2(\alpha-\delta)}{\textrm{c}^3}\cos\varphi -
\frac{|\boldsymbol{b}||\boldsymbol{V}|^2 (\beta -\delta)}{\textrm{c}^3}\cos\varphi\cos^2\theta \\
& + \frac{|\boldsymbol{b}||\boldsymbol{V}|^2(\beta-\delta-1)}{\textrm{c}^3}  \sin\theta \cos\theta \sin\varphi \cos A\\
= & \frac{|\boldsymbol{b}||\boldsymbol{V}|^2(\alpha-\beta)}{\textrm{c}^3}\cos\varphi +
\frac{|\boldsymbol{b}||\boldsymbol{V}|^2 (\beta -\delta)}{\textrm{c}^3}\cos\varphi\sin^2\theta \\
& + \frac{|\boldsymbol{b}||\boldsymbol{V}|^2(\beta-\delta-1)}{\textrm{c}^3}  \sin\theta \cos\theta \sin\varphi \cos A
\end{aligned}
\end{equation}
manifests itself as a combination of the Kennedy-Thorndike experiment testing the factor $(\alpha-\beta+1)$ and the Michelson-Morley experiment testing the factor $(\beta+\delta-\frac{1}{2})$ \citep{Michelson87, Kennedy32, Mansouri77b}. In special relativity, the factors equal to $(\alpha-\beta) = 0$ and $(\beta-\delta) = \frac{1}{2}$, therefore only the two last terms are included in the Equation~(\ref{groupdelay_gcrs}) describing the group delay. The difference $(\beta-\delta)$ is tested twice, once with the geodetic parameters (second term in the Equation~(\ref{geomdel_s_2terms})), and another time with the apparent displacement of the extragalactic radio sources (third term in the Equation~(\ref{geomdel_s_2terms})). Thus, the second order aberration effect in the source positions is an alteration of the Michelson-Morley experiment, which was already noted in~\citet{Klioner12}.\\
If $(\alpha-\beta) \neq 0$ in the Equation~(\ref{geomdel_s_2terms}), the Earth scale factor estimated from geodetic VLBI data would differ from unity by the factor $\frac{|\boldsymbol{V}|^2 (\alpha - \beta)}{\textrm{c}^2}$. In this context, the discrepancy between the VLBI and the SLR (Satellite Laser Ranging) scale factor ($1.37\pm0.10$~ppb), reported by~\citet{Altamimi16} for the International Terrestrial Reference Frame ITRF2014, may be interpreted as the violation of the local Lorenz invariance at the level of $(\alpha-\beta) \approx 0.14\pm0.01$. However, the modern laboratory Kennedy-Thorndike tests of the Lorenz invariance (e.g., \citet{Herrmann09}) rule this interpretation out.

All three terms in the Equation~(\ref{geomdel_s_2terms}) can be presented in the form of Equation~(\ref{dtau1}), i.e., as a combination of estimated parameters and partial derivatives. Therefore, the corrections $\Delta b$ and $\Delta s$ are given by the following equations (note, that the angle $A$ shows the difference of directions from the source to vectors $\boldsymbol{b}$ and $\boldsymbol{V}$, and therefore the function $\cos A$ is a part of the partial derivative):
\begin{equation}\label{delta_bs}
\begin{aligned}
\Delta b &= -\frac{|\boldsymbol{b}||\boldsymbol{V}|^2(\alpha-\beta)}{\textrm{c}^2} -
\frac{|\boldsymbol{b}||\boldsymbol{V}|^2 (\beta -\delta)}{\textrm{c}^2}\sin^2\theta \\
\Delta s &= \frac{|\boldsymbol{V}|^2(\beta-\delta-1)}{\textrm{c}^2}  \sin\theta \cos\theta .
\end{aligned}
\end{equation}
The barycentric velocity $\boldsymbol{V}$ is well-known from the high-precision Solar system ephemerids. Therefore, the Equation~(\ref{delta_bs}) is a standard part of the delay model, i.e., Equation~(\ref{groupdelay_gcrs}).\\
Let's apply the Equation~(\ref{delta_bs}) to a hypothetical preferred reference frame. In this preferred reference frame (e.g., Cosmic Microwave Background (CMB)) the direction of the velocity vector $\boldsymbol{V}$ is constant. Therefore, if $(\beta-\delta) \neq \frac{1}{2}$ the shift in source positions depends on the angle $\theta$ only, i.e., the systematic shift across the sky would show a quadrupole pattern.

If we define the vector $\boldsymbol{s} = (\cos \hat{\alpha} \cos \hat{\delta}, \sin \hat{\alpha} \cos \hat{\delta},\\ \sin \hat{\delta})$ where $\hat{\alpha}$ and $\hat{\delta}$ are the right ascension and declination of the source, respectively, and the vector $\boldsymbol{V} = (V_x, V_y, V_z)$, we can further reformulate the Equation~(\ref{delta_bs}) for $\Delta s$ using the relationship $|\boldsymbol{V}| \cos\theta = (\boldsymbol{s}\cdot\boldsymbol{V})$, the spherical law of cosine for the angle $\theta$, and the division of $\sin \theta$ in two equations which gives us the corresponding corrections to the source coordinates $\Delta \hat{\alpha} \cos \hat{\delta}$ and $\Delta \hat{\delta}$ as
\begin{equation}\label{RA}
\begin{aligned}
\Delta \hat{\alpha} \cos \hat{\delta} = &\frac{(\beta-\delta-1)}{\textrm{c}^2} \bigg( \Big(V^2_y - V^2_x\Big) \cos \hat{\delta} \sin \hat{\alpha} \cos \hat{\alpha} \\
&+ V_x V_y \cos 2\hat{\alpha} \cos \hat{\delta} - V_x V_z \sin \hat{\alpha} \sin \hat{\delta} \\
&+ V_y V_z \cos \hat{\alpha} \sin \hat{\delta} \bigg)
\end{aligned}
\end{equation}
and
\begin{equation}\label{De}
\begin{aligned}
\Delta \hat{\delta} = &\frac{(\beta-\delta-1)}{\textrm{c}^2} \Bigg( -\frac{1}{2} V^2_x \sin 2\hat{\delta} \cos^2 \hat{\alpha} -  \frac{1}{2} V^2_y \sin 2\hat{\delta} \sin^2 \hat{\alpha} \\
&+\frac{1}{2} V^2_z \sin 2\hat{\delta} - \frac{1}{2} V_x V_y \sin 2\hat{\delta} \sin 2\hat{\alpha} \\
&+ V_x V_z \cos 2\hat{\delta} \cos \hat{\alpha}
+ V_y V_z \cos 2\hat{\delta} \sin \hat{\alpha}
\Bigg) .
\end{aligned}
\end{equation}
A similar equation is developed for the scale factor from the $\Delta b$ in the Equation~(\ref{delta_bs}) using the formula $\sin^2 \theta = 1 - \cos^2 \theta$ in addition:
\begin{equation}\label{b}
\begin{aligned}
\Delta b =& -\frac{|\boldsymbol{b}||\boldsymbol{V}|^2(\alpha-\beta)}{\textrm{c}^2} -
\frac{|\boldsymbol{b}|(\beta -\delta)}{\textrm{c}^2} \\
&\Bigg(-V^2_x \cos^2 \hat{\alpha} \cos^2 \hat{\delta} -V^2_y \sin^2 \hat{\alpha} \cos^2 \hat{\delta}  \\
&-V^2_z \sin^2 \hat{\delta} - V_x V_y \cos^2 \hat{\delta} \sin 2\hat{\alpha} \\
&- \frac{1}{2} V_x V_z \cos \hat{\alpha} \sin 2\hat{\delta}- \frac{1}{2} V_y V_z \sin \hat{\alpha} \sin 2\hat{\delta}
\Bigg)
\end{aligned}
\end{equation}
which means that the scale factor magnitude depends on the equatorial coordinates of the observed radio sources.

\section{Methodology comments}
Additional consideration of the special relativity in the geodetic VLBI may be developed in a form of an arbitrary synchronization of clock instead of the Einstein synchronization adopted in Equation~(\ref{xt}) or equivalent to the introduction of the preferred reference frame as in~\citet{Klioner12}.

Any non-Einstein synchronisation comes down to replacement of the velocity $\boldsymbol{V}$ in the second equation of the Equation~(\ref{xt}) by a function $\varepsilon(\boldsymbol{V}) \neq \boldsymbol{V}$. In this sense, the geodetic velocity of the second station $\boldsymbol{w}_{2}$, so far ignored throughout this paper, could be explicitly introduced to modify the equation for the time transformation in Equation~(\ref{xt}) as follows:
\begin{equation}\label{xt2}
t' = A\Bigg(t - \frac{\boldsymbol{V}+\boldsymbol{w}_{2}}{\textrm{c}^2}\boldsymbol{x}\Bigg) .
\end{equation}
This modification immediately results in the appearance of additional terms in equation of the relativistic time delay:
\begin{equation}\label{d_tau_geom}
\Delta\tau_{g}=-\frac{(\boldsymbol{b}\cdot\boldsymbol{s})}{\textrm{c}} \cdot \frac{(\boldsymbol{w}_{2}\cdot\boldsymbol{s})}{\textrm{c}}
\end{equation}
as a part of Equation~(\ref{groupdelay_gcrs}) and in addition to the Equation~(\ref{groupdelay_gcrs_simple}). This correction changes the VLBI scale factor only and it is not relevant to astrometric positions of reference radio sources, therefore, it is not part of the final equation by~\citet{Klioner12}. Further analysis of this term is essential because it lies besides the traditional discussion on the clock synchronization (see, for example, discussion by~\citet{Cole76}).

Therefore, there is no need in the introduction of the preferred reference frame explicitly in the VLBI relativistic group delay equation. The barycentric and geocentric celestial reference systems (BCRS and GCRS) are fully sufficient to test the Lorentz invariance. The fact of the matter is that GCRS is moving with respect to BCRS along with the Earth. Therefore, direction of the relative velocity of GCRS with respect to BCRS changes as time progresses. This effectively introduces a set of different inertial reference frames along the Earth's orbit. So, introduction of BCRS and GCRS does not mean that we deal with only two frames. Instead of that an infinite number of the inertial frames along the Earth's orbit are introduced. Therefore, tracking the consistency of VLBI time-delay residuals over one orbital revolution allows to compare VLBI observations conducted at different time of year from different locations of GCRS moving in various directions and, thus, to test Lorentz invariance.

\section{Conclusions}
We can conclude that a variety of opportunities is allowed by the geodetic VLBI technique to test the Lorenz invariance in a frame of the kinematic RMS formalism. However, precision of the ground based VLBI measurements is not competitive to the laboratory experiments. While the geodetic VLBI is able to reach an accuracy of the estimation of the $\alpha, \beta$ and $\delta$ parameters at the level of $\sim10^{-2}$ using the barycentric velocity of the Earth in approximation \citep{Smoot77}, the laboratory tests set bounds on the anisotropy of the speed of light to $\sim10^{-12}$ with the Michelson-Morley experiments \citep{Herrmann09} and to $\sim10^{-8}$ with the Kennedy-Thorndike experiments \citep{Tobar10} using the velocity of the Sun with respect to the CMB which is about $\sim 370$~km/s \citep{Smoot77}. Theoretically, space VLBI observations within, e.g., the RadioAstron mission at baselines $\sim50$ times longer than the Earth radius reduced to the CMB reference frame (as proposed by~\citet{Klioner12}) may provide an accuracy of $\sim10^{-6}$ for the $\alpha, \beta$ and $\delta$ parameter combinations.

\begin{acknowledgements}
The authors thank the anonymous reviewers for their suggestions and comments which helped to improve the manuscript significantly. We acknowledge the IVS and all its components for providing VLBI data \citep{Nothnagel15}. Hana Kr{\'a}sn{\'a} works within the Hertha Firnberg position T697-N29, funded by the Austrian Science Fund (FWF). This paper has been published with the permission of the Geoscience Australia CEO.
\end{acknowledgements}

\bibliographystyle{spbasic} % style aa.bst
\bibliography{references_TitovKrasna_IAG17} % your references Yourfile.bib

\end{document}